\documentclass[journal]{IEEEtran}
\usepackage{cite}
\usepackage{textcomp}
\usepackage{graphicx}
\graphicspath{{./}}
\usepackage{booktabs,tabularx,array}
\usepackage{colortbl}
\usepackage{xcolor}
\usepackage[hidelinks]{hyperref}

\definecolor{ink}{HTML}{18334A}
\definecolor{teal}{HTML}{147D83}
\definecolor{pale}{HTML}{EDF5F6}
\definecolor{amber}{HTML}{A36520}

\newcolumntype{Y}{>{\raggedright\arraybackslash}X}

\title{Sustaining In-Network Computation in Satellite Networks: An Architecture for Semantic Continuity}

\author{
Zhiyuan Ren,~\IEEEmembership{Member,~IEEE,}~Wenchi Cheng,~\IEEEmembership{Senior Member,~IEEE}
\thanks{This work was supported by the National Key Research and Development Program of China (No. 2024YFE0200302).}%
\thanks{Zhiyuan Ren and Wenchi Cheng are with the School of Telecommunications Engineering, Xidian University, Xi'an 710071, China.}%
\thanks{Corresponding author: Zhiyuan Ren (zyren@xidian.edu.cn).}}

\begin{document}
\bstctlcite{IEEEexample:BSTcontrol}
\raggedbottom
\setlength{\dbltextfloatsep}{8pt plus 2pt minus 2pt}
\setlength{\textfloatsep}{8pt plus 2pt minus 2pt}
\setlength{\floatsep}{8pt plus 2pt minus 2pt}
\maketitle

\begin{abstract}
Satellite computations often outlast the paths and providers initially serving them. We argue that in-network computing should organize changing execution opportunities around one continuing logical computation: workers execute ready operations, while consumers recognize which results already contribute. We present an architecture separating logical computation, stream context, and physical binding, and show in wildfire assessment how retained observations support recomputation while reachable association history stays in place. A 300-node Linux prototype examines placement, dependency assembly, and continuation. Across 120 images, stage-wise placement delivers 82.08\% of positions within deadlines versus 76.06\% for single-node placement, while traces separate blocked output delivery from dependencies split across providers. Across 50 observations with 20-node isolation, Confirmation-Coordinated Rebinding (CCR) coordinates replanning, source replay, cross-attempt acceptance, and in-band confirmation to complete all 50 observations with verified outputs using seven replays and 13 ignored duplicate outputs.
\end{abstract}


\section{Introduction}

Satellite monitoring increasingly requires processing observations, associating them with earlier evidence, and delivering timely reports. Onboard filtering on missions such as Phi-Sat-1 \cite{phisat} and orbital edge computing \cite{orbitaledge} supply onboard compute, yet a continuing assessment spans many observations while serving satellites and paths change. Execution therefore depends on input access and downstream reachability alongside processor availability.

In low Earth orbit (LEO), changing inter-satellite links (ISLs) and ground contacts can prevent an operational processor from obtaining inputs or delivering results in time, whereas another satellite holds needed data, compatible compute, and a usable onward path. These conditions define an execution opportunity. Because inputs can be retained across contacts, receiving, computing, and delivering may occupy separate intervals; the challenge is organizing transient opportunities into one continuing computation.

\begin{figure}[!t]
\centering
\includegraphics[width=\columnwidth]{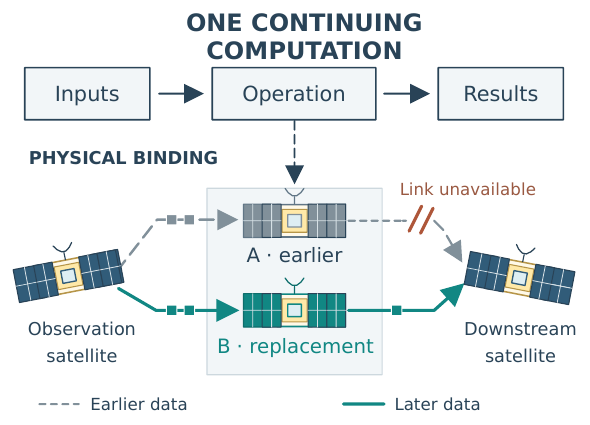}
\caption{One logical computation uses changing providers and paths. A and B perform the same operation at different times; earlier data can remain in flight as later work reaches B. The logical operation persists while its physical execution changes.}
\label{fig:concept}
\end{figure}

In-network computing (INC) places computation along transmission paths through programmable network devices \cite{sincn}, as in industrial edge systems \cite{incmobile} and holographic service slices \cite{incslices}. Participating nodes process task data and return outputs to forwarding once a required input group is ready: selection terminates payloads, aggregation combines inputs, and transformation creates new representations. Processing changes what travels next, while connectivity governs where processing can occur. Before a brief downlink, compact outputs may fit remaining capacity when raw observations cannot; an along-path node finishing that reduction in time supplies an opportunity unavailable at the source.

We argue that INC should be a basic execution form in satellite computing, supporting along-path processing as communication needs evolve. Sustaining execution across provider changes requires keeping accepted contributions, in-flight results, and unfinished input groups associated with one logical work (Fig.~\ref{fig:concept}). We call this preservation of operation requirements, dependencies, and result interpretation \emph{semantic continuity}.

Related work addresses computing-dependent multipath routing \cite{cdmr}, integrated sensing, communication, and computing \cite{satelliteisc}, and cooperative space--ground inference \cite{leoedge}. Stream-processing and network-function systems provide proven building blocks: MillWheel maintains persistent state and exact-once record delivery across worker changes over stable storage \cite{millwheel}, while OpenNF coordinates network-function state transfers with controller-driven forwarding updates \cite{opennf}. In dynamic satellite paths, however, along-path computation alters data representations while transient contacts shift where execution can proceed. Sustaining satellite INC therefore adds a distinct architectural responsibility: jointly maintaining operation dependencies, accepted partial contributions, and onward forwarding choices as both data representations and execution opportunities evolve.

This article makes three contributions:
\begin{itemize}
 \item An architectural case for sustaining in-network computation across changing providers and paths around one continuing logical work.
 \item A three-part architecture separating logical computation, stream context, and physical binding, illustrated through a multi-observation wildfire assessment design.
 \item Packet-path evidence comparing placement, diagnosing split dependencies, and tracing verified continuation through source replay, cross-attempt result acceptance, and in-band confirmation.
\end{itemize}

\section{Execution Opportunities and Continuity Requirements}

A \emph{continuing logical stream} identifies a task, its operations, and input--output relationships for continuous monitoring or bounded jobs. Data units---messages, sensor records, tensor slices, or image tiles---keep their logical relationships while execution moves among physical nodes. Within such a stream, a useful execution opportunity combines access to required inputs and state, compatible processing, and an onward path meeting task deadlines. Because filtering leaves selected records for a contact whereas feature extraction leaves compact vectors, a candidate's value depends on what is already computed and what must still travel. Independent work units often finish within one assignment before later units use another provider, whereas a mid-operation change requires accounting for received inputs and in-flight outputs. Both cases occur within one task, so continued execution combines waiting, rerouting, and selective recomputation, finishing when remaining dependencies, resources, and contacts meet the deadline.

Packetization sets transport boundaries, whereas each operation defines the input group a worker must assemble---from a single record to a spatial neighborhood or complete object. Three dependency patterns govern continuation across provider changes: independent units restart from one retained input; spatial or multi-branch stages assemble neighboring or matching inputs released once dependent outputs finish; and temporal association relies on historical state across reporting intervals. Changing a provider, forwarding path, or access attachment during an active task is \emph{rebinding}. Redirecting later inputs can divide an unfinished group between workers; continuation therefore brings those dependencies together from retained state or reconstructed inputs before the deadline expires.

Semantic continuity keeps successive executions within the same logical work through three requirements:
\begin{enumerate}
 \item \emph{Compatible execution}: replacements satisfy declared input formats, preprocessing, model or operator versions, and numerical tolerances.
 \item \emph{Available dependencies}: required inputs and retained state can be assembled before the next operation executes.
 \item \emph{Interpretable outcomes}: downstream stages distinguish accepted results, repeated execution, valid empty results, and unfinished work.
\end{enumerate}

Under these rules, an accepted detection can advance an assessment while other regions remain unresolved, keeping result validity, progress, and task completion distinct as tasks specify whether reports may include gaps or require complete coverage. Computed outputs need onward delivery, whereas unfinished operations need missing inputs or history. These requirements map directly to three architectural responsibilities: logical computation defines acceptable execution and results; stream context relates data to ongoing work; and physical binding connects providers with inputs and consumers. While runtime-programmable switches support packet-consistent partial reconfiguration \cite{runtime}, continued task execution further preserves operation, dependency, and result relationships across physical updates.

\section{Architecture and Execution Workflow}

\begin{figure*}[!t]
\centering
\includegraphics[width=0.98\textwidth]{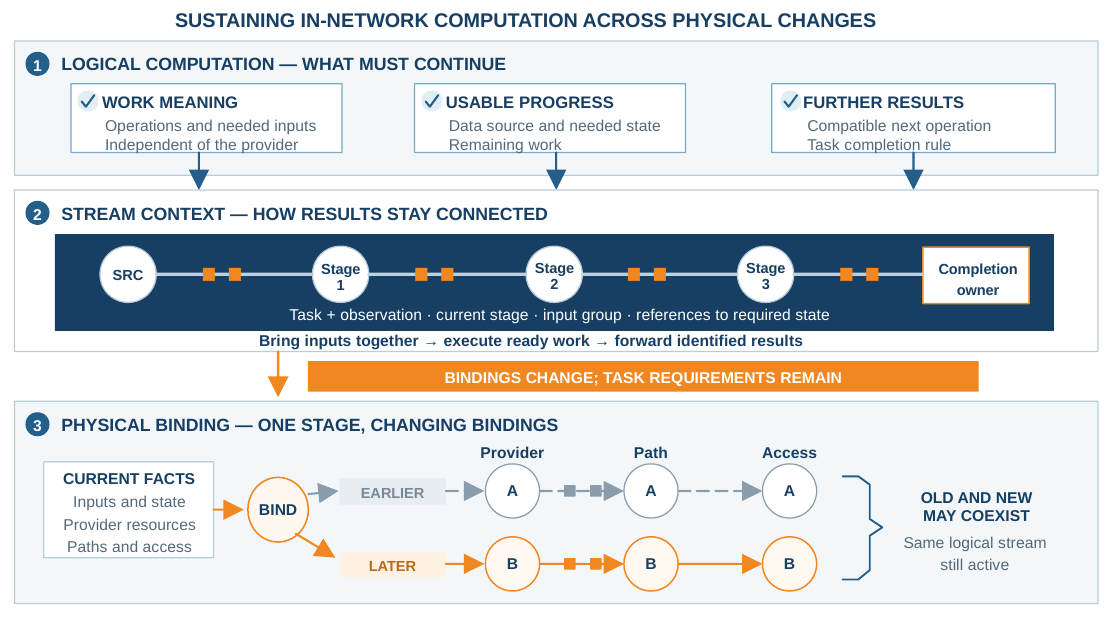}
\caption{Three responsibilities organize continued execution. Logical computation defines acceptable work. Stream context connects data to that work. Physical binding chooses providers and paths; workers execute when the required inputs and state are available. Results return to the network with their updated context.}
\label{fig:architecture}
\end{figure*}

The architecture in Fig.~\ref{fig:architecture} organizes local processing around continuing work: applications describe the computation, node-local workers execute ready operations, binding logic selects providers and paths, and designated receivers (completion owners) apply result-acceptance rules.

\subsection{Three Architectural Responsibilities}\label{sec:archresp}

\emph{Logical computation and task descriptions.} A versioned task description shared by binding logic, workers, and receivers specifies operations, dependencies, acceptance rules, expected coverage, and reporting deadlines. Each stage declares input and output formats, model or algorithm versions, and numerical tolerances; inference stages specify tensor shapes, normalization, weights, and thresholds, while temporal association specifies required history. These declarations govern progress reuse---retained inputs regenerate results, intermediate representations feed downstream stages, and accumulated history stays accessible or transfers---while coverage rules distinguish complete reports from reports with gaps.

\emph{Stream context and execution workers.} A node-local worker uses stream context to associate arriving data with its task, stage, and input group, buffering dependencies, invoking the operator once ready, and returning outputs to forwarding with updated metadata so context identifies what belongs together while the network delivers or recovers inputs. Coordinates identify spatial neighbors, branch identifiers align converging inputs, and sequence metadata links observations to history. Persistent logical work identifiers and per-attempt result-set identifiers travel in headers or reference task descriptions so replacements interpret arriving data after upstream changes. In spatial filtering, coordinates track missing neighbors per acquisition and emit each completed neighborhood immediately, while deadlines and acceptance records keep qualified earlier results eligible for unaccepted work.

\emph{Physical binding and forwarding.} Physical binding selects compatible instances and routes connecting each provider to its dependencies and output consumer. Stages can share one satellite to avoid intermediate transfers or span satellites along the path, guided by representation size: an intermediate tensor can exceed its source observation or final detection records. Forwarding directs arriving data to the selected local worker---via user-space sockets or onboard accelerators---and carries outputs onward. Route updates can retain that worker, whereas provider changes select another compatible instance; if deadlines permit, data can wait for a later contact. In the prototype, distributed topology and capability facts support node forwarding while the ingress computes task plans (Section~\ref{sec:binding}).

\begin{figure*}[!t]
\centering
\includegraphics[width=0.98\textwidth]{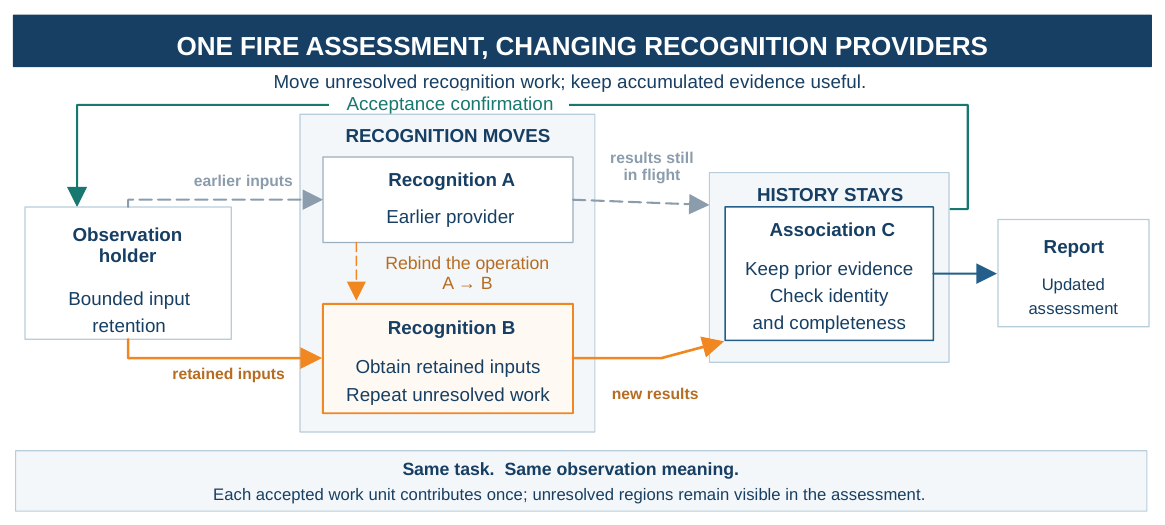}
\caption{Design example: recognition moves while evidence association stays in place. Retained observations enable B to recompute unresolved work. Satellite C returns an acceptance confirmation for each accepted work unit, authorizing the observation holder to release buffered source inputs. Qualified earlier results from A remain eligible for unresolved work within its deadline. Arrows represent logical data relationships and may traverse several network hops.}
\label{fig:story}
\end{figure*}

\subsection{Task Setup, Execution, and Rebinding Workflow}\label{sec:workflow}

During \emph{setup}, the source submits a processing graph and destination to its ingress, where running operator instances advertise capabilities; the ingress selects instances, resolves the destination's access node, and installs an execution policy before transmission. During \emph{execution}, workers assemble and process ready input groups, returning outputs with updated context; downstream recognition can start on a reduced group while other groups are still arriving, and the receiver tracks accepted contributions against reporting rules.

During \emph{rebinding}, updated connectivity or capability facts prompt a replacement plan: the ingress switches to a surviving precomputed backup chain or recomputes a chain over surviving links. Using explicit hop-by-hop Segment Routing over IPv6 (SRv6) \texttt{End.X} adjacency segments, the prototype reconciles task policies independently of batch route installation and redirects subsequent traffic. Replay regenerates unfinished work from retained inputs while the receiver preserves accepted contributions across attempts; source-received confirmation releases retained inputs and cancels pending retries.

\section{Use Case: Continuous Wildfire Assessment}

Wildfire assessment must incorporate new observations while preserving useful historical evidence. Timely delivery under disrupted infrastructure and limited resources is also central to emergency networking \cite{uavemergency}. In Fig.~\ref{fig:story}, recognition extracts candidate hotspots from multispectral tiles, association combines detections with historical fire perimeters, and reporting produces periodic assessments. The task describes the observation area and expected tiles. An observation holder, such as the imaging satellite, retains source inputs within a storage budget and retention deadline.

\subsection{Selective Rebinding and Source Replay}\label{sec:wfreplay}

Recognition initially runs at A, while C maintains association history. When A's input path becomes unavailable, C remains reachable through an alternate route. The ingress selects B with compatible model weights, available capacity, and paths delivering inputs and results before the reporting deadline, advancing recognition while preserving the assessment accumulated at C. Because tiles are recognized independently, once the replacement path installs, the holder replays unconfirmed tiles so B recomputes recognition without fetching A's intermediate features, and B's outputs route to C to continue association against existing history without state migration.

Different forms of progress thus use different continuation methods: retained observations regenerate recognition results, and reachable association state preserves earlier contributions. The holder's retention budget bounds how far back recognition can be retried; an evicted observation leaves unresolved work if no usable result arrives. Whereas the Section~\ref{sec:proto} prototype keeps one fixed endpoint receiver across changing access satellites, relocating an orbiting completion owner such as C is an architectural extension requiring a consistent checkpoint of association state and the accepted Work-ID ledger, with ownership handoff preventing concurrent commits.

\subsection{Result-Set Acceptance, Retention, and Reporting}\label{sec:wfaccept}

A \emph{Work ID} combines the task, observation epoch, tile coordinates, and model version, while each attempt has a separate \emph{Result-Set ID}. Records within a set carry item identifiers, and a completion record states the expected count and payload digest, including count zero for a no-hotspot outcome. Satellite C assembles records only within one Result Set and counts retransmitted items once; it never merges partial sets from A and B.

The task accepts the first complete qualified set. Satellite C persists the accepted set and its contribution before confirming acceptance to the holder; later sets for that Work ID trigger only a repeat confirmation. Results from A stay eligible while unaccepted and within deadline. If A and B each emit two candidate records, receiving one from each leaves two incomplete sets; C waits for both records and the completion record from one attempt, whereas a count-zero completion record confirms a valid no-hotspot tile.

Acceptance confirmation allows early release of the corresponding source tile. If feedback is delayed or lost, the holder retains the tile until its local retention deadline or storage limit requires eviction. Eviction ends that replay opportunity without changing C's recorded acceptance decision, as the holder's lack of confirmation and the receiver's lack of a result are different states. Finally, C groups accepted detections by source observation time: at a reporting deadline, the report uses accepted evidence for that interval, enters a completed no-hotspot result as a negative finding, and marks unresolved tiles as ``Unassessed Data Gaps,'' bringing contributions from different satellites into one assessment under unified reporting rules.

\section{Binding and Deployment Choices}\label{sec:binding}

\begin{table}[!t]
\caption{Provider Choice and the Coordination It Requires}
\label{tab:binding}
\centering\small
\renewcommand{\arraystretch}{1.1}
\begin{tabularx}{\columnwidth}{@{}>{\raggedright\arraybackslash}p{0.22\columnwidth}Y Y@{}}
\toprule
\rowcolor{ink}
\color{white}\textbf{Decision} & \color{white}\textbf{Ingress binding} & \color{white}\textbf{Network-resolved binding}\\
\midrule
Choose providers & At ingress, for the complete plan & Near each stage's execution\\
\rowcolor{pale}
Use available knowledge & Coordinate stage placement and connecting paths & Adapt using information available along the path\\
Change the plan & Replace the plan for subsequent traffic & Change one stage's provider\\
\rowcolor{pale}
Keep related inputs together & Route them to one provider; plan state handoff & Account for where inputs and state are held\\
\bottomrule
\end{tabularx}
\end{table}

Table~\ref{tab:binding} compares two ways to select physical execution opportunities. With \emph{ingress binding}, topology and capability facts support a joint choice of operator instances and connecting paths, carried by native SRv6 in the prototype; when a plan becomes unusable, the ingress updates subsequent traffic while in-flight packets keep their earlier plan. With \emph{network-resolved binding}, the segment list names logical services that forwarding nodes resolve to providers from neighbor, topology, and capability facts, allowing physical choices to change near execution while the logical sequence stays fixed provided the chosen worker obtains required inputs and state. Joint planning is useful when dependent stages should share a node or avoid a costly intermediate transfer, whereas local resolution is useful when several compatible instances offer changing opportunities along the path, with both arrangements following the same task requirements.

Execution and onward transmission must also fit available contacts and orbital power constraints: a fast processor is ineffective if its outlet closes first, whereas a slower node with a timely contact still meets the deadline. Predicted contacts guide model pre-positioning and planned handoffs, while unscheduled disruptions depend on remaining inputs, paths, and replanning latency. Sunlight and eclipse also change energy availability \cite{phoenix}. Because processing and transmission share a satellite's power budget, retaining data for a later contact or choosing another provider can outperform immediate local execution, much as satellite--terrestrial quality-of-service (QoS) provisioning weighs freshness alongside delay and reliability \cite{satelliteaoi} so delayed work still contributes if it arrives before the reporting deadline.

The next operation further determines which retained representation is useful. Because intermediate activations can exceed the size of compressed observations, transferring the observation and recomputing may use fewer bytes, whereas an available intermediate result avoids repeating earlier stages and compact detection records can advance association even though they cannot reconstruct feature extraction. Choosing among these possibilities joins execution progress with the remaining transmission budget. Overall communication cost includes payloads, encapsulation, context, recovery traffic, and traversed links, where batching amortizes headers at the cost of delaying a ready group and compact references reduce packet size while requiring accessible descriptions.

\section{Packet-Path Prototype and Observations}\label{sec:proto}

Three experiments examine ingress-bound execution of a spatial pipeline across placement, split dependencies, and continuation. Together they evaluate timely delivery over 120 images, trace how mid-flight rebinding divides unconfirmed inputs, and test coordinated recovery across 50 observations.

\subsection{Prototype Setup and Spatial Pipeline}\label{sec:protosetup}

The 300-node platform comprises 200 satellite network nodes and 100 endpoints in isolated network and process environments on one Linux host. Native SRv6 uses explicit hop-by-hop \texttt{End.X} adjacency segments to connect selected operators and destination access nodes. Operator interfaces use the Linux eXpress Data Path (XDP) and \texttt{AF\_XDP} sockets for node-local user-space processing \cite{xdp}, returning outputs to forwarding.

A separate host-side \texttt{AF\_XDP} fabric emulates 1~Mb/s links with 4~MiB drop-tail queues. Propagation delays follow geometric link distances at the speed of light across 550-km polar and 600-km inclined orbits. A fixed source endpoint (node~141) transmits to a fixed destination endpoint (node~240) whose satellite access attachment changes over time, keeping the accepted-position ledger at one receiver without state migration. In Confirmation-Coordinated Rebinding (CCR), confirmations travel in-band on UDP port~46202 through a reverse SRv6 policy over surviving links while global IPv6 Forwarding Information Base (FIB) updates converge: once all 40 positions are accepted, the receiver reassembles the frame, verifies its SHA-256 digest against the deterministic reference, logs completion, and sends a three-packet confirmation burst (at 0, 0.4, and 1.5~s) against reverse-path loss (plus one repeat per late duplicate output). Source receipt releases retained inputs and cancels retries.

The pipeline converts a reference 240\texttimes160 satellite RGB image, sent as a 15\texttimes10 grid of 150 tiles at 10-ms intervals, to grayscale; downsamples it to 120\texttimes80 pixels; applies a 3\texttimes3 mean filter; and performs tile-local Sobel edge detection. Under the fixed-receiver simplification noted in Section~\ref{sec:wfreplay}, this four-stage pipeline directly exercises spatial neighborhood assembly (in 3\texttimes3 filtering) and deterministic position-level Result-Set acceptance. Downsampling yields 40 output positions in an 8\texttimes5 grid of 16\texttimes16 tiles; the final column has eight valid and eight zero-padded pixel columns. The blur worker waits for neighboring boundary samples, while UDP stream context carries work identifiers, attempt identifiers, and spatial coordinates.

Along this pipeline, in-network computation sharply reduces downstream transmission demand. In observation~2 of the continuation run, payload volume drops from 124,800 bytes at the source to 48,000 after grayscale, 12,800 after downsampling, and 3,840 after edge detection (forty 96-byte packets: 64-byte stream context plus a 32-byte 16\texttimes16 binary edge mask), excluding outer UDP/IPv6/SRv6 headers. In a separate filtering test, a classifier forwards 14 of 150 source tiles at its deployed threshold and terminates the remaining 136 (90.7\%) at the operator, illustrating traffic selection by computation.

\subsection{Placement Delivery and Provider-Change Diagnostics}\label{sec:protoplace}

Paired 120-image runs use the same implementation, orbital scene, seeded fault schedule, and operator deployment. Because all 300 nodes share one host, the planner assumes homogeneous node capacity with stage compute weights of 1, 4, 9, and 2 (1, 4, 9, and 2~ms estimated stage times), advertised memory gates of 64, 256, 576, and 64~KiB across grayscale, downsampling, 3\texttimes3 filtering, and Sobel stages, and stage transfer volumes over 1~Mb/s links, while native C \texttt{AF\_XDP} workers run unthrottled at sub-millisecond host speed---isolating communication-bound payload reduction while keeping compute-weight trade-offs active in planning. Single-node execution places all four stages on one provider; stage-wise execution selects providers per stage (co-locating when beneficial) from estimated compute and transfer completion times.

Admitting 120 back-to-back images at 10-ms tile intervals (4,800 expected positions, 60-s deadlines, 20 satellites isolated at 120~s), single-node execution delivers 3,651 timely valid positions (76.06\%, 90 complete images) versus 3,940 (82.08\%, 98 complete images) for stage-wise execution, with 2,856 and 3,048 positions delivered before isolation. Under continuous admission near the 1~Mb/s link rate, early along-path grayscale and downsampling shrink payloads from 124.8~KB to 12.8~KB before downstream hops, relieving link serialization and 4~MiB queue pressure.

To isolate how a mid-image provider change affects unconfirmed work, a diagnostic trace without replay follows acquisitions 261--267 across rebinding, counting intermediate positions after 3\texttimes3 filtering and final positions after Sobel processing. Acquisition 261 delivers all 40 final positions; in acquisitions 262 and 263, the initial provider completes all 40 intermediate positions, but downstream interruption limits final delivery to 14 and zero positions. During acquisition 264, installing a new plan while 150 source tiles are in flight splits the 40 positions across three outcomes: the initial worker completes 13 intermediate positions that stall at a downstream relay; the replacement completes nine positions reaching the receiver; and 18 fail to assemble because their 3\texttimes3 boundary neighbors divide between providers, expiring after five idle seconds. Acquisitions 265--267 deliver all 40 final positions along the new path while acquisition 264 stays incomplete, separating blocked onward delivery from split neighborhood assembly.

\begin{figure}[!t]
\centering
\includegraphics[width=\columnwidth]{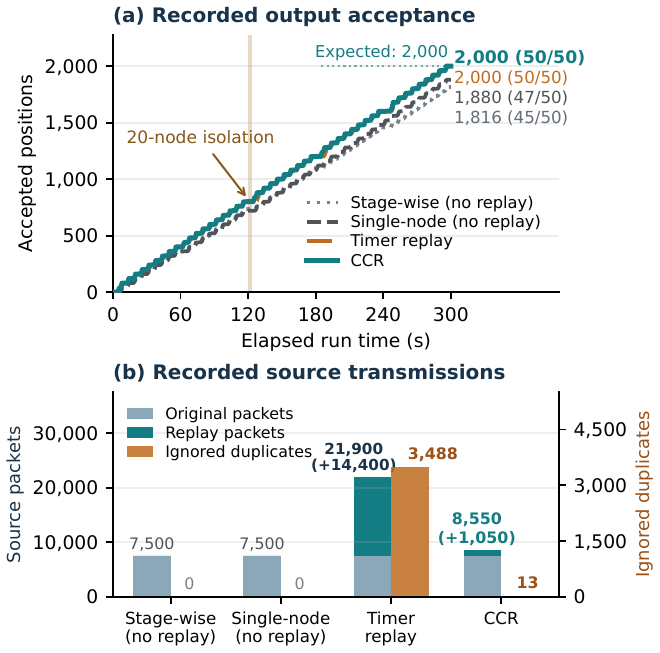}
\caption{Matched 50-observation executions on the same prototype implementation, orbital scene, and fault schedule, with 20-node isolation at 120~s: stage-wise and single-node execution without replay, unconfirmed timer replay, and CCR. (a) Cumulative accepted output positions. (b) Original and replay source packets, with ignored duplicate-output counts.}
\label{fig:execution-comparison}
\end{figure}

\subsection{Continuing Earlier Work through Replay}\label{sec:protoreplay}

Figure~\ref{fig:execution-comparison} compares four matched 50-observation executions admitted every 6~s starting around 0.2--0.3~s (0.31~s in CCR) over a 300-s window, with 20-node fabric isolation at 120.0~s. All four runs share the same implementation, independent task-policy reconciliation, orbital scene, fault schedule, and operator deployment, comparing no-replay placement, unconfirmed timer replay, and CCR's coordinated continuation policy.

Without replay, stage-wise execution completes 45 observations (1,816 positions), leaving observations~1, 21, 31, and~41 at zero positions and observation~11 at 16 after mid-emission access rebinding. Single-node execution completes 47 observations (1,880 positions), losing only~1, 11, and~21 because co-locating all four operators avoids the inter-stage ISLs interrupted during observations~31 and~41: under sparse 6-s admission where 1~Mb/s links stay uncongested, multi-hop ISL exposure outweighs the upstream payload reduction that prevails under continuous admission (Section~\ref{sec:protoplace}), making semantic continuation essential.

Under ingress binding, in-flight packets keep their earlier segment list. Rather than buffering outputs stranded downstream (Section~\ref{sec:protoplace}) or resolving providers along the path (Section~\ref{sec:binding}), the prototype retains each observation's 150 source tiles for 30~s (up to two retries) and replays from the source. Downsampling and 3\texttimes3 filtering couple tiles across output positions, so byte-stream retransmission cannot directly map missing outputs to single source packets. Because all four integer operators are deterministic, each position forms a single-item Result Set requiring no separate completion record, letting the receiver preserve accepted positions across attempts. Collecting all 40 positions triggers frame reassembly and SHA-256 verification against the deterministic reference (filling the count-and-digest check role of Section~\ref{sec:wfaccept}) before the receiver logs completion and emits its confirmation burst to release buffered source tiles. Frame-level confirmation avoids reverse bitmaps here, whereas partial-coverage tasks or larger frames can return selective bitmap confirmations under those acceptance rules.

Both replay modes arm an 8-s retry timer after admission (and 8~s after Attempt~1 ends) and mark unconfirmed retained observations due when a new binding installs. Without confirmation, timer replay buffers completed observations for 30~s; rebinding marks them due ahead of newly interrupted work in first-in, first-out order, creating head-of-line replay contention and blind retries. CCR closes this loop: confirmation evicts completed observations, withholds Attempt~2 unless a newer binding installs (or retention nears expiry), and schedules a 1.5-s follow-up if rebinding occurs mid-retry.

\begin{figure}[!t]
\centering
\includegraphics[width=\columnwidth]{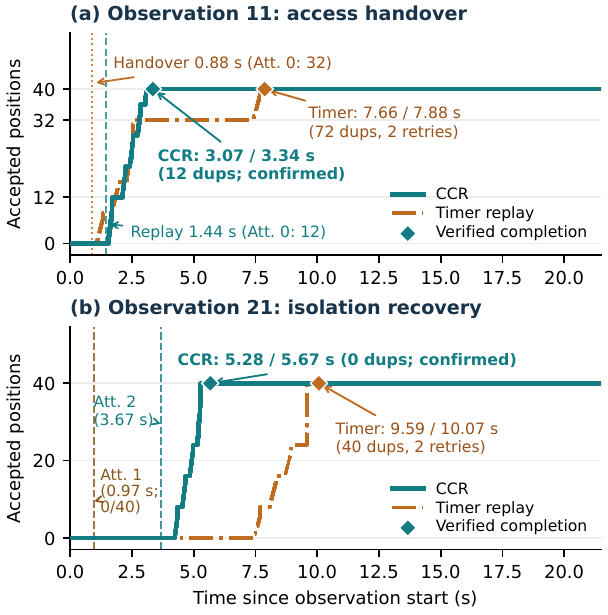}
\caption{Position acceptance and verified completion in unconfirmed timer replay and CCR on the same prototype implementation. (a) Observation~11: timer replay accepts all 40 positions by 7.66~s and verifies completion at 7.88~s; CCR reaches these milestones at 3.07 and 3.34~s, then sends its confirmation burst. (b) Observation~21: timer replay reaches them at 9.59 and 10.07~s after queued retries of earlier observations, whereas CCR reaches them at 5.28 and 5.67~s with zero duplicate outputs. Times are relative to each observation's start.}
\label{fig:continuation}
\end{figure}

Figure~\ref{fig:continuation}(a) compares observation~11 across access handover. In timer replay, handover at 0.88~s steers trailing Attempt-0 tiles to deliver 32 positions, but replaying already completed observations~8--10 delays Attempt~1 until 6.35~s; all 40 positions are accepted by 7.66~s and verified at 7.88~s, followed by a blind second retry that brings ignored duplicates to 72. In CCR, prior confirmations have evicted observations~1--10; rebinding at 1.44~s starts Attempt~1 immediately while trailing Attempt-0 tiles deliver 12 positions and Attempt~1 supplies the remaining 28, reaching 40 accepted positions at 3.07~s and verified confirmation at 3.34~s with 12 duplicates and Attempt~2 canceled.

Figure~\ref{fig:continuation}(b) tracks observation~21, admitted just after the 120.0-s isolation event. In timer replay, Attempt~0 delivers zero positions and queued retries of completed observations~18--20 delay Attempt~1, reaching 40 accepted positions at 9.59~s and verified completion at 10.07~s before a second retry adds 40 duplicates. In CCR, an initial access handover starts Attempt~1 at 0.97~s, which enters still-converging post-isolation topology under the prototype's 1-s neighbor keepalive (3-miss timeout) and flooded link-state updates: local neighbor timeout first installs an interim route before flooded updates expose a downstream break, leaving Attempt~1 with zero outputs. Installing a converged replacement chain at 3.67~s triggers Attempt~2, completing 40 accepted positions by 5.28~s and SHA-256 verification and confirmation at 5.67~s with zero duplicates, while subsequent observations resume single-attempt completion.

Both timer replay and CCR complete all 50 observations (2,000 positions), whereas unconfirmed timer replay executes 96 replays in window versus seven under CCR's coordinated confirmation, buffer-eviction, retry-gating, and rebinding-follow-up policy across observations~1, 11, 21, 31, and~41 (one on~1, 11, 31; two on~21, 41). For the 40 observations whose 60-s deadlines fall within the 300-s window (observations~1--40; observations~41--50 have deadlines past 300~s), deadline completions are 36, 37, 40, and~40 across stage-wise, single-node, timer replay, and CCR. Source packets total 7,500 (stage-wise), 7,500 (single-node), 21,900 (timer replay: 14,400 replay packets, 3,488 ignored duplicates), and 8,550 (CCR: 1,050 replay packets---a 92.7\% reduction from the coordinated policy---and 13 duplicates, 12 on observation~11 and one on~41), with all 163 confirmations (26-byte UDP payloads: three per observation plus 13 duplicate-triggered repeats) reaching the source across four reverse-policy updates.

\section{Research Challenges and Open Issues}

Scaling semantic continuity across operational constellations raises three open research challenges. Addressing them requires balancing storage budgets, metadata synchronization, and execution determinism under intermittent connectivity.

\emph{Progress retention and replay scheduling.} Which information must stay available between execution opportunities? Source inputs permit recomputation, intermediate results avoid completed stages, and accepted contributions preserve accumulated outcomes, with each item's value depending on the next operation and where compute and contacts emerge. Retention policies must weigh storage held by earlier work against new inputs and deadline feasibility. An execution policy must also decide when to compute, wait, or move data, and when several unfinished observations overlap in flight, order replays by deadline and cap replay rate so recovery does not delay new admissions. Useful evaluation connects retained bytes, repeated computation, and transmission cost to completed work, while recovery for temporal association must preserve acceptance decisions alongside evidence so replay order does not alter committed contributions.

\emph{Context distribution and state availability.} An available processor needs task descriptions, models, and calibration data to become useful to ongoing work. Compact identifiers reduce packet overhead but create dependencies on accessible descriptions, so their placement should follow likely execution opportunities while accounting for synchronization traffic and fetch latency. Because stable model descriptions and changing execution records have different update needs, descriptions can be pre-positioned before predicted contacts while compact work identifiers accompany data. An open question is how a node obtains enough current state to act accurately when task progress spans several nodes, involving both representation and timely access beyond header size.

\emph{Compatibility across heterogeneous providers.} Successive providers may use different preprocessing, FP32 or INT8 precision, model versions, and accelerator backends. Compatibility descriptions must expose the properties needed by the next stage, while task rules specify acceptable output variation. Deterministic selection is needed when qualified attempts must produce the same accepted outcome across runs; whereas the evaluated operators produce identical outputs to permit position-level assembly across attempts, tasks using heterogeneous implementations must define numerical and spatial consistency requirements and the scope at which a complete Result Set is accepted. A practical approach separates validating an implementation's operating envelope from per-assignment checks: the former establishes formats and numerical behavior, while the latter verifies that those properties and current resources satisfy the task, letting heterogeneous nodes participate under consistent result semantics.

\section{Conclusion}

Satellite networks can organize processing along changing paths around one continuing logical computation. The proposed architecture connects local execution opportunities through operation requirements, stream context, and physical binding, while the wildfire design shows how retained inputs and accepted evidence support subsequent work. Packet-path experiments relate placement to delivery, expose split dependencies, and show earlier observations continuing through replay while preserving accepted contributions. In CCR, source replay and cross-attempt position acceptance complete all 50 observations with verified outputs, while in-band confirmations reach the source for every observation and release retained inputs. Sustaining in-network computation requires these relationships to persist so changing satellites contribute useful progress to the same task.

\bibliographystyle{IEEEtran}
\bibliography{references_space}

\end{document}